\documentclass[11pt]{iopart}

\newcommand{\bea}{\begin{eqnarray}}
\newcommand{\eea}{\end{eqnarray}}

\begin{document}

\title{Newtonian limit of fully nonlinear cosmological perturbations in Einstein's gravity}
\author{Jai-chan Hwang}
\address{Department of Astronomy and Atmospheric Sciences,
         Kyungpook National University, Daegu 702-701, Republic of Korea}
\ead{jchan@knu.ac.kr}
\author{Hyerim Noh}
\address{Korea Astronomy and Space Science Institute,
         Daejeon 305-348, Republic of Korea}
\ead{hr@kasi.re.kr}


\begin{abstract}

We prove that in the infinite speed-of-light limit
(i.e., non-relativistic and subhorizon limits),
the relativistic fully nonlinear cosmological perturbation
equations in two gauge conditions, the zero-shear gauge and
the uniform-expansion gauge, exactly reproduce the Newtonian
hydrodynamic perturbation equations in the cosmological background;
as a consequence, in the same two gauge conditions, the Newtonian
hydrodynamic equations are exactly recovered in the Minkowsky background.

\end{abstract}




%
%
\section{Introduction}

Newton's gravity and Einstein's gravity are two competing and
complementary theoretical frames where the current cosmological
research is practically based. Only with the advent of Einstein's
gravity the history of modern physical cosmology has begun
\cite{Einstein-1917}. Equations describing the cosmological world
model were first derived based on Einstein's gravity by Friedmann in
1922 \cite{Friedmann-1922}, and the Newtonian study followed later
by Milne and McCrea in 1934 \cite{Milne-McCrea-1934}. The two
results coincided in the zero-pressure limit. Similarly, the linear
perturbation equations were first derived based on Einstein's
gravity by Lifshitz in 1946 \cite{Lifshitz-1946}, and the Newtonian
study followed later by Bonnor in 1957 \cite{Bonnor-1957}. For
density perturbation, the two results again coincided in the
zero-pressure limit \cite{Nariai-1969}.

The Newtonian limit of Einstein's gravity in cosmology, however, is
not a settled subject. The problem is rather serious in the level of
background world model where the well known Newtonian cosmology
\cite{Milne-McCrea-1934} is known to be a special one motivated and
guided by the results in Einstein's gravity; without additional
symmetry in the matter distribution or the boundary condition at
infinity the Newtonian dynamics and gravitation could lead to the
world models different from the Friedmann's one based on Einstein's
gravity, see \cite{Layzer-1954}. Even in the post-Newtonian approach
in cosmology we have to {\it subtract} the background equations
based on Einstein's gravity to get proper Newtonian equations for
the perturbation \cite{Hwang-etal-2008}.

Despite such a trouble in the background level, curiously we
often have quite successful relativistic/Newtonian correspondence
in the perturbation level. Exact Newtonian perturbation equations
are recovered in the Newtonian limit of the post-Newtonian
approximation independently of the gauge condition \cite{Hwang-etal-2008};
exact Newtonian equations are recovered in Minkowsky background
\cite{Chandrasekhar-1965}. In the perturbation theory, however,
the analysis and results crucially depend on the gauge choice
\cite{Bardeen-1980,Hwang-1994,Hwang-Noh-1999}.

The exact correspondence of all three Newtonian perturbation
variables (perturbed density, velocity, and gravitational potential)
is not available in a single gauge condition even to the linear order
\cite{Hwang-1994,Hwang-Noh-1999}. Such a correspondence of all
variables is available only in the subhorizon limit in the
zero-shear gauge and the uniform-expansion gauge \cite{Hwang-Noh-1999},
which is shown to be valid even to the second order in perturbation
\cite{Hwang-etal-2012}.

However, whether the relativistic/Newtonian correspondence in the
subhorizon limit is valid to the fully nonlinear order was not known
in the literature. Here we provide a proof of the correspondence in
the two fundamental gauge conditions. For the correspondence we will
show that we need weak gravity and slow-motion limits, in addition
to the subhorizon limit as well as negligible pressure and internal
energy density compared with the energy density; all these limits
can be summarized as the infinite speed-of-light limit. Such a proof
is now available mainly due to a recent advent of the exact and
fully nonlinear cosmological perturbation theory in Einstein's
gravity \cite{Hwang-Noh-2013}; for the concrete nature of our formulation with a certain limit caused by {\it ignoring} the transverse-tracefree part of the metric tensor, see section \ref{sec:NL-eqs}.

Section \ref{sec:NL-eqs} is a summary of the fully nonlinear
cosmological perturbation theory presented in \cite{Hwang-Noh-2013}.
Section \ref{sec:correspondence} presents our main result,
proving the Newtonian limit.
In sections \ref{sec:E-conservation} and \ref{sec:Newtonian-limit} we clarify a couple of issues. Section \ref{sec:Discussion} is a discussion.

%
%
%
\section{Fully nonlinear perturbations}
                                             \label{sec:NL-eqs}

We consider the scalar- and vector-type perturbations in
a {\it flat} background with the metric convention
\cite{Bardeen-1988,Hwang-Noh-2013}
\bea
   & & ds^2 = - a^2 \left( 1 + 2 \alpha \right) d x^0 d x^0
       - 2 a \chi_i d x^0 d x^i
       + a^2 \left( 1 + 2 \varphi \right) \delta_{ij} d x^i d x^j,
   \label{metric-PT}
\eea where $a(x^0)$ is the cosmic scale factor, and $\alpha$,
$\varphi$ and $\chi_i$ are functions of spacetime with {\it
arbitrary} amplitudes; index of $\chi_i$ is raised and lowered by
$\delta_{ij}$ as the metric; notice that $x^0 = \eta$ with $a d \eta
\equiv c dt$. The spatial part of the metric is simple because we
already have taken the spatial gauge condition without losing any
generality to fully nonlinear order
\cite{Bardeen-1988,Hwang-Noh-2013}.

The arbitrary perturbation variables $\alpha$, $\varphi$ and
$\chi_i$ give only five independent degrees of freedom whereas we
should have six independent physical degrees of freedom for the most
general perturbations. This is because we have {\it ignored} the
transverse-tracefree part corresponding to two degrees of freedom
associated with the gravitational waves (to the linear order) and other transverse-tracefree distortion of the three-dimensional hypersurface to the nonlinear order. In addition, although we
have fixed the spatial gauge condition with three gauge (congruence)
degrees of freedom, we have {\it not} fixed the temporal gauge
(slicing) degree of freedom yet, thus enabling the basic set of
equations presented in a sort of gauge-ready form. Thus, our
perturbation variables contain four physical degrees of freedom (two
for scalar-type perturbation and the other two for the vector-type
perturbation) and one temporal gauge degree of freedom. We have
freedom to impose the slicing condition depending on the
mathematical simplification and/or physical interpretation. Our
equations are designed so that after imposing any of the fundamental
slicing condition, the gauge degrees of freedom are completely
removed, and consequently each variable in those gauge can be
equivalently regarded as a unique gauge-invariant variable, to
nonlinear perturbation order, see equation
(\ref{temporal-gauges-NL}) and below.

In order to have the fully nonlinear and exact equations it is
essentially important to {\it take} the spatial gauge condition as
we do and {\it ignore} the transverse-tracefree part of the metric.
By taking the first condition we do not lose any generality
especially because the spatial gauge condition we take is the unique
one which (together with any temporal gauge condition in the pool of
our suggested fundamental slicing conditions) leaves the remaining
variables spatially gauge invariant. Whereas neglecting the
transverse-tracefree part of the metric should be regarded as an
important shortcoming of our formulation. In this sense our formulation can be regarded as {\it not} exact. At the moment we cannot
formulate the fully nonlinear and exact equations in the presence of
the transverse-tracefree part of the metric, and these should be
handled perturbatively only. We still call our formulation fully
nonlinear and exact despite this important shortcoming because we
have {\it not} imposed any condition on our metric and
energy-momentum perturbation variables and formulated equations in
exact forms, see equations (\ref{eq1})-(\ref{K-bar-eq}).

We consider a fluid without anisotropic stress.
The energy momentum tensor is given as
\bea
   \widetilde T_{ab} = \widetilde \varrho c^2 \widetilde u_a \widetilde u_b
       + \widetilde p \left( \widetilde g_{ab}
       + \widetilde u_a \widetilde u_b \right),
   \label{Tab}
\eea where tildes indicate covariant quantities; $\widetilde u_a$ is
the normalized fluid four-vector; $\widetilde \varrho$ and
$\widetilde p$ are the mass density and pressure, respectively. We
decompose the fluid quantities to the background and perturbation as
\bea
   \widetilde \varrho = \varrho + \delta \varrho, \quad
       \widetilde p = p + \delta p, \quad
       \widetilde u_i \equiv a {v_i \over c},
\eea where the index of $v_i$ is raised and lowered by $\delta_{ij}$
as the metric. In the explicit presence of the internal energy
$\widetilde \varrho \widetilde \Pi$, $\widetilde \varrho$ should be
replaced by $\widetilde \varrho (1 + \widetilde \Pi/c^2)$
\cite{Chandrasekhar-1965}; in the latter expression $\widetilde
\varrho$ is the material density.

In \cite{Hwang-Noh-2013} we have introduced several different
definitions of the fluid three-velocity (see the Appendix D in
\cite{Hwang-Noh-2013}). Although mathematically equivalent, in this
work we will use the following definition. The fluid three-velocity
measured by the Eulerian observer with the normal four-vector
$\widetilde n^c$ is introduced as \bea
   & & \widehat V^i
       \equiv {\widetilde h^{(n)i}_{\;\;\;\;\; c} \widetilde u^c
       \over - \widetilde n_c \widetilde u^c}
       = {1 \over N} \left( {\widetilde u^i \over \widetilde u^0} + N^i \right),
\eea where $\widetilde h^{(n)}_{ab} \equiv \widetilde g_{ab} +
\widetilde n_a \widetilde n_b$ is the projection tensor normal to
$\widetilde n^c$, and the index of $\widehat V^i$ is raised and
lowered by the ADM three-space metric $h_{ij}$; $N$ and $N_i$ are
the lapse and shift vector, respectively, in the ADM formulation. In
order to use the perturbation notation, we introduce \bea
   & & \widehat V_i \equiv a {\widehat v_i \over c},
\eea where the index of $\widehat v_i$ is raised and lowered by
$\delta_{ij}$. Compared with $v_i$ we have \bea
   & & v_i \equiv \widehat \gamma \widehat v_i,
\eea where \bea
   & & \widehat \gamma
       = \sqrt{ 1 + {v^k v_k \over c^2 (1 + 2 \varphi)} }
       = {1 \over \sqrt{ 1
       - {\widehat v^k \widehat v_k \over c^2 (1 + 2 \varphi)}}},
\eea is the Lorentz factor. In terms of $\widehat v_i$ the Lorentz
factor becomes a well known form.

We can decompose $\chi_i$, $v_i$ and $\widehat v_i$ to scalar- and
vector-type perturbations even to the nonlinear order as \bea
   \chi_i = c \chi_{,i} + a \Psi_i^{(v)}, \quad
       v_i \equiv - v_{,i} + v_i^{(v)}, \quad
       \widehat v_i \equiv - \widehat v_{,i} + \widehat v^{(v)}_i,
\eea with $\Psi^{(v)i}_{\;\;\;\;\;\;,i} \equiv 0$ and
$v^{(v)i}_{\;\;\;\;\;\;,i} \equiv 0 \equiv \widehat
v^{(v)i}_{\;\;\;\;\;\;,i}$. Due to the nonlinear relation between
$v_i$ and $\widehat v_i$ the scalar- and vector-decompositions for
$v_i$ and $\widehat v_i$ do not coincide with each other to the
nonlinear order. To the nonlinear order {\it our} scalar- and
vector-type perturbations are coupled in the equation level.

We assign dimensions as \bea
   & & [a] = [\widetilde g_{ab}] = [\widetilde u_a]
       = [\alpha] = [\varphi] = [\chi^i] = [v^i/c] = [\widehat v^i/c] = 1, \quad
       [v/c] = L,
   \nonumber \\
   & &
       [x^i] = L, \quad
       [\chi] = T, \quad
       [\kappa] = T^{-1}, \quad
       [\widetilde T_{ab}] = [\widetilde \varrho c^2]
       = [\widetilde p], \quad
       [G \widetilde \varrho] = T^{-2},
\eea where $\kappa$, a perturbed part of the trace of extrinsic
curvature (equivalently, a perturbed part of the expansion scalar
of the normal-frame vector with a minus sign) will be introduced in equation (\ref{eq1}).

Now, using $\widehat v_i$ as the fluid three-velocity, a complete
set of fully nonlinear perturbation equations without taking the
temporal gauge are the following \cite{Hwang-Noh-2013}.

\noindent
Definition of $\kappa$: \bea
   \fl \kappa
       \equiv
       3 {\dot a \over a} \left( 1 - {1 \over {\cal N}} \right)
       - {1 \over {\cal N} (1 + 2 \varphi)}
       \left[ 3 \dot \varphi
       + {c \over a^2} \left( \chi^k_{\;\;,k}
       + {\chi^{k} \varphi_{,k} \over 1 + 2 \varphi} \right)
       \right].
   \label{eq1}
\eea
ADM energy constraint:
\bea
   \fl - {3 \over 2} \left( {\dot a^2 \over a^2}
       - {8 \pi G \over 3} \widetilde \varrho
       - {\Lambda c^2 \over 3} \right)
       + {\dot a \over a} \kappa
       + {c^2 \Delta \varphi \over a^2 (1 + 2 \varphi)^2}
   \nonumber \\
   \fl \qquad
       = {1 \over 6} \kappa^2
       - 4 \pi G \left( \widetilde \varrho + {\widetilde p \over c^2} \right)
       \left( \widehat \gamma^2 - 1 \right)
       + {3 \over 2} {c^2 \varphi^{,i} \varphi_{,i} \over a^2 (1 + 2 \varphi)^3}
       - {c^2 \over 4} \overline{K}^i_j \overline{K}^j_i.
   \label{eq2}
\eea
ADM momentum constraint:
\bea
   \fl {2 \over 3} \kappa_{,i}
       + {c \over 2 a^2 {\cal N} ( 1 + 2 \varphi )}
       \left( \Delta \chi_i
       + {1 \over 3} \chi^k_{\;\;,ik} \right)
       + 8 \pi G \left( \widetilde \varrho + {\widetilde p \over c^2} \right)
       a \widehat \gamma^2 {\widehat v_{i} \over c^2}
   \nonumber \\
   \fl \qquad
       =
       {c \over a^2 {\cal N} ( 1 + 2 \varphi)}
       \Bigg\{
       \left( {{\cal N}_{,j} \over {\cal N}}
       - {\varphi_{,j} \over 1 + 2 \varphi} \right)
       \left[ {1 \over 2} \left( \chi^{j}_{\;\;,i} + \chi_i^{\;,j} \right)
       - {1 \over 3} \delta^j_i \chi^k_{\;\;,k} \right]
   \nonumber \\
   \fl \qquad
       - {\varphi^{,j} \over (1 + 2 \varphi)^2}
       \left( \chi_{i} \varphi_{,j}
       + {1 \over 3} \chi_{j} \varphi_{,i} \right)
       + {{\cal N} \over 1 + 2 \varphi} \nabla_j
       \left[ {1 \over {\cal N}} \left(
       \chi^{j} \varphi_{,i}
       + \chi_{i} \varphi^{,j}
       - {2 \over 3} \delta^j_i \chi^{k} \varphi_{,k} \right) \right]
       \Bigg\}.
   \label{eq3}
\eea
Trace of ADM propagation:
\bea
   \fl - 3 {1 \over {\cal N}}
       \left( {\dot a \over a} \right)^{\displaystyle\cdot}
       - 3 {\dot a^2 \over a^2}
       - 4 \pi G \left( \widetilde \varrho + 3 {\widetilde p \over c^2} \right)
       + \Lambda c^2
       + {1 \over {\cal {\cal N}}} \dot \kappa
       + 2 {\dot a \over a} \kappa
       + {c^2 \Delta {\cal N} \over a^2 {\cal N} (1 + 2 \varphi)}
   \nonumber \\
   \fl \qquad
       = {1 \over 3} \kappa^2
       + 8 \pi G \left( \widetilde \varrho + {\widetilde p \over c^2} \right)
       \left( \widehat \gamma^2 - 1 \right)
       - {c \over a^2 {\cal N} (1 + 2 \varphi)} \left(
       \chi^{i} \kappa_{,i}
       + c {\varphi^{,i} {\cal N}_{,i} \over 1 + 2 \varphi} \right)
       + c^2 \overline{K}^i_j \overline{K}^j_i.
   \label{eq4}
\eea
Tracefree ADM propagation:
\bea
   \fl \left( {1 \over {\cal N}} {\partial \over \partial t}
       + 3 {\dot a \over a}
       - \kappa
       + {c \chi^{k} \over a^2 {\cal N} (1 + 2 \varphi)} \nabla_k \right)
       \Bigg\{ {c \over a^2 {\cal N} (1 + 2 \varphi)}
   \nonumber \\
   \fl \qquad
       \times
       \left[
       {1 \over 2} \left( \chi^i_{\;\;,j} + \chi_j^{\;,i} \right)
       - {1 \over 3} \delta^i_j \chi^k_{\;\;,k}
       - {1 \over 1 + 2 \varphi} \left( \chi^{i} \varphi_{,j}
       + \chi_{j} \varphi^{,i}
       - {2 \over 3} \delta^i_j \chi^{k} \varphi_{,k} \right)
       \right] \Bigg\}
   \nonumber \\
   \fl \qquad
       - {c^2 \over a^2 ( 1 + 2 \varphi)}
       \left[ {1 \over 1 + 2 \varphi}
       \left( \nabla^i \nabla_j - {1 \over 3} \delta^i_j \Delta \right) \varphi
       + {1 \over {\cal N}}
       \left( \nabla^i \nabla_j - {1 \over 3} \delta^i_j \Delta \right) {\cal N} \right]
   \nonumber \\
   \fl \qquad
       =
       8 \pi G \left( \widetilde \varrho + {\widetilde p \over c^2} \right)
       \left[ {\widehat \gamma^2 \widehat v^i \widehat v_j \over c^2 (1 + 2 \varphi)}
       - {1 \over 3} \delta^i_j \left( \widehat \gamma^2 - 1 \right)
       \right]
       + {c^2 \over a^4 {\cal N}^2 (1 + 2 \varphi)^2}
   \nonumber \\
   \fl \qquad
       \times
       \Bigg[
       {1 \over 2} \left( \chi^{i,k} \chi_{j,k}
       - \chi_{k,j} \chi^{k,i} \right)
       + {1 \over 1 + 2 \varphi} \left(
       \chi^{k,i} \chi_k \varphi_{,j}
       - \chi^{i,k} \chi_j \varphi_{,k}
       + \chi_{k,j} \chi^k \varphi^{,i}
       - \chi_{j,k} \chi^i \varphi^{,k} \right)
   \nonumber \\
   \fl \qquad
       + {2 \over (1 + 2 \varphi)^2} \left(
       \chi^{i} \chi_{j} \varphi^{,k} \varphi_{,k}
       - \chi^{k} \chi_{k} \varphi^{,i} \varphi_{,j} \right) \Bigg]
       - {c^2 \over a^2 (1 + 2 \varphi)^2}
   \nonumber \\
   \fl \qquad
       \times
       \Bigg[ {3 \over 1 + 2 \varphi}
       \left( \varphi^{,i} \varphi_{,j}
       - {1 \over 3} \delta^i_j \varphi^{,k} \varphi_{,k} \right)
       + {1 \over {\cal N}} \left(
       \varphi^{,i} {\cal N}_{,j}
       + \varphi_{,j} {\cal N}^{,i}
       - {2 \over 3} \delta^i_j \varphi^{,k} {\cal N}_{,k} \right) \Bigg].
   \label{eq5}
\eea
Covariant energy conservation:
\bea
   \fl
       \left[ {\partial \over \partial t}
       + {1 \over a ( 1 + 2 \varphi )} \left( {\cal N} \widehat v^k
       + {c \over a} \chi^k \right) \nabla_k \right] \widetilde \varrho
       + \left( \widetilde \varrho + {\widetilde p \over c^2} \right)
       \Bigg\{
       {\cal N} \left( 3 {\dot a \over a} - \kappa \right)
   \nonumber \\
   \fl \qquad
       +
       {({\cal N} \widehat v^k)_{,k} \over a (1 + 2 \varphi)}
       + {{\cal N} \widehat v^k \varphi_{,k} \over a (1 + 2 \varphi)^2}
       + {1 \over \widehat \gamma}
       \left[ {\partial \over \partial t}
       + {1 \over a ( 1 + 2 \varphi )} \left( {\cal N} \widehat v^k
       + {c \over a} \chi^k \right) \nabla_k \right] \widehat \gamma \Bigg\}
       = 0.
   \label{eq6}
\eea
Covariant momentum conservation:
\bea
   \fl {1 \over a \widehat \gamma}
       \left[ {\partial \over \partial t}
       + {1 \over a ( 1 + 2 \varphi )} \left( {\cal N} \widehat v^k
       + {c \over a} \chi^k \right) \nabla_k \right]
       \left( a \widehat \gamma \widehat v_i \right)
       + \widehat v^k \nabla_i \left( {c \chi_k \over a^2 ( 1 + 2
       \varphi)} \right)
   \nonumber \\
   \fl \qquad
       + {c^2 \over a} {\cal N}_{,i}
       - \left( 1 - {1 \over \widehat \gamma^2} \right) {c^2 {\cal N}
       \varphi_{,i} \over a (1 + 2 \varphi)}
   \nonumber \\
   \fl \qquad
       + {1 \over \widetilde \varrho + {\widetilde p \over c^2}}
       \left\{
       {{\cal N} \over a \widehat \gamma^2} \widetilde p_{,i}
       + {\widehat v_i \over c^2}
       \left[ {\partial \over \partial t}
       + {1 \over a ( 1 + 2 \varphi )} \left( {\cal N} \widehat v^k
       + {c \over a} \chi^k \right) \nabla_k \right] \widetilde p \right\}
       = 0,
   \label{eq7}
\eea
where \bea
   \fl {\cal N} \equiv \sqrt{ 1 + 2 \alpha
       + {\chi^k \chi_k \over a^2 ( 1 + 2 \varphi )}}, \quad
       \overline{K}^i_j \overline{K}^j_i
       = {1 \over a^4 {\cal N}^2 (1 + 2 \varphi)^2}
       \Bigg\{
       {1 \over 2} \chi^{i,j} \left( \chi_{i,j} + \chi_{j,i} \right)
       - {1 \over 3} \chi^i_{\;\;,i} \chi^j_{\;\;,j}
   \nonumber \\
   \fl - {4 \over 1 + 2 \varphi} \left[
       {1 \over 2} \chi^i \varphi^{,j} \left(
       \chi_{i,j} + \chi_{j,i} \right)
       - {1 \over 3} \chi^i_{\;\;,i} \chi^j \varphi_{,j} \right]
       + {2 \over (1 + 2 \varphi)^2} \left(
       \chi^{i} \chi_{i} \varphi^{,j} \varphi_{,j}
       + {1 \over 3} \chi^i \chi^j \varphi_{,i} \varphi_{,j} \right) \Bigg\},
   \nonumber \\
   \label{K-bar-eq}
\eea with ${\cal N}$ introduced as $N \equiv a {\cal N}$; $N$ is the ADM lapse function with $N \equiv 1/\sqrt{-\widetilde g^{00}}$. These equations were derived in Section 3 of
\cite{Hwang-Noh-2013}; here we express the equations using $\widehat
v_i$ instead of $v_i$ used in \cite{Hwang-Noh-2013}, set the energy
density as $\widetilde \mu \equiv \widetilde \varrho c^2$, and
recover $c$; in our notation $\widetilde \varrho$ includes the
internal energy.

To the background order, equations (\ref{eq2}), (\ref{eq4})
and (\ref{eq6}) give \bea
   \fl
       {\dot a^2 \over a^2} = {8 \pi G \over 3} \varrho
       + {\Lambda c^2 \over 3}, \quad
       {\ddot a \over a} = - {4 \pi G \over 3}
       \left( \varrho + 3 {p \over c^2} \right) + {\Lambda c^2 \over 3}, \quad
       \dot \varrho + 3 {\dot a \over a}
       \left( \varrho + {p \over c^2} \right) = 0.
   \label{BG-equations}
\eea As mentioned, contrary to common beliefs, even with
$p = 0 = \Lambda$, these equations are pure relativistic in their origin;
for a clarification, see \cite{Layzer-1954}.

Equations (\ref{eq1})-(\ref{eq7}) are presented without taking the
temporal gauge (hypersurface or slicing) condition. As the temporal
gauge condition we can impose any one of the following conditions
\bea
   & & {\rm comoving \; gauge:}              \hskip 2.33cm     \widehat v \equiv 0,
   \nonumber \\
   & & {\rm zero\!-\!shear \; gauge:}        \hskip 1.94cm     \chi \equiv 0,
   \nonumber \\
   & & {\rm uniform\!-\!curvature \; gauge:} \hskip .50cm      \varphi \equiv 0,
   \nonumber \\
   & & {\rm uniform\!-\!expansion \; gauge:} \hskip .47cm      \kappa \equiv 0,
   \nonumber \\
   & & {\rm uniform\!-\!density \; gauge:}   \hskip 1.00cm      \delta \equiv 0,
   \label{temporal-gauges-NL}
\eea or combinations of these to each perturbation order; we may
call these the fundamental gauge conditions. With the imposition of
any of these temporal gauge conditions the remaining perturbation
variables are free from the remnant (spatial and temporal) gauge
mode, and have unique gauge-invariant combinations. Thus, we can
regard each perturbation variable in these gauges as the
gauge-invariant one to the nonlinear order
\cite{Bardeen-1988,Hwang-Noh-2013}. The temporal synchronous gauge
condition sets $\alpha = 0$; however, as this condition fails to fix
the gauge degree of freedom completely, and leaves the remnant gauge
mode even from the linear order, we do not consider it as the
fundamental gauge condition in the nonlinear perturbation analysis. As the comoving gauge condition we suggested $\widehat v \equiv 0$ which is the same as $v \equiv 0$ for vanishing vector-type perturbation; in the presence of vector-type perturbation these two conditions differ from each other from the third-order perturbation.

%
%
%
\section{Newtonian correspondence}
                                          \label{sec:correspondence}

%
%
\subsection{Newtonian equations}
                                          \label{sec:Newtonian-eqs}

The Newtonian hydrodynamic equations in the cosmological background are \bea
   & & \dot {\widetilde \varrho}
       + 3 {\dot a \over a} \widetilde \varrho
       = - {1 \over a} \nabla \cdot
       \left( \widetilde \varrho {\bf v} \right),
   \label{mass-conservation} \\
   & & \dot {\bf v} + {\dot a \over a} {\bf v}
       + {1 \over a} {\bf v} \cdot \nabla {\bf v}
       = {1 \over a} \nabla U
       - {1 \over a \widetilde \varrho} \nabla \widetilde p,
   \label{momentum-conservation} \\
   & & {\Delta \over a^2} U
       = - 4 \pi G \left( \widetilde \varrho - \varrho \right),
   \label{Poisson-eq}
\eea where $\widetilde \varrho$ is the material density. These
equations properly reduce to the Newtonian hydrodynamic equations in
the Minkowsky background where the background order quantities
become $a \equiv 1$ and $\varrho \equiv0$ with no $\Lambda$. From
equations (\ref{mass-conservation}) and (\ref{Poisson-eq}) we have
\bea
   & & \dot U + {\dot a \over a} U
       = 4 \pi G a \Delta^{-1}
       \nabla \cdot \left( \widetilde \varrho {\bf v} \right).
   \label{U-eq}
\eea Our aim in this work is to show that equations
(\ref{eq1})-(\ref{eq7}) properly reduce to equations
(\ref{mass-conservation})-(\ref{U-eq}) in the infinite
speed-of-light limit in the zero-shear gauge and in the
uniform-expansion gauge to the fully nonlinear order.

Equation (\ref{mass-conservation}) can be decomposed to the
background order part (the third one of equation
(\ref{BG-equations}) with vanishing pressure term) and the perturbed
part \bea
   & & \dot \delta = - {1 \over a} \nabla \cdot
       \left[ \left( 1 + \delta \right) {\bf v} \right],
   \label{mass-conservation-pert}
\eea where $\delta \equiv \delta \varrho/\varrho$. Newtonian
derivation of equations (\ref{momentum-conservation}),
(\ref{Poisson-eq}) and (\ref{mass-conservation-pert}) can be found
in sections 7-9 of \cite{Peebles-1980}.

%
%
\subsection{Infinite speed-of-light limit}
                                          \label{sec:limit}

The relativistic perturbation variables ($\alpha$, $\varphi$, $\widehat v^i$,
$\widetilde \varrho$ and $\widetilde p$) in this paragraph are valid
in both the zero-shear gauge and the uniform-expansion gauge. As the
non-relativistic limit we consider \bea
   \alpha \ll 1, \quad
       \varphi \ll 1, \quad
       \widehat v^k \widehat v_k / c^2 \ll 1, \quad
       \widetilde p \ll \widetilde \varrho c^2, \quad
       \widetilde \Pi/c^2 \ll 1.
   \label{NL-limit}
\eea
We identify \bea
   \alpha = - {1 \over c^2} U, \quad
       \varphi = {1 \over c^2} V, \quad
       v^i = \widehat v^i = {\bf v},
   \label{identification}
\eea where ${\bf v}$ is the perturbed Newtonian velocity; $U$ and
$V$ correspond to the Newtonian and the post-Newtonian perturbed
gravitational potentials, respectively
\cite{Chandrasekhar-1965,Hwang-etal-2008}; later we will show
$\varphi = - \alpha$, thus $V = U$. As the subhorizon limit, we take
the dimensionless quantity \bea
   {k^2 c^2 \over a^2 H^2} \gg 1,
   \label{SS-limit}
\eea where $k$ the comoving wave-number with $\Delta = - k^2$ in the
Fourier space notation; $H \equiv \dot a/a$; in the presence of
$\Lambda$ we consider $H^2 \sim 8 \pi G \varrho$. The first two
conditions in equation (\ref{NL-limit}) are the weak gravity
(general relativistic) limits, the third one is the slow-motion
(special relativistic) limit, and the last two conditions imply that we
ignore the pressure and internal energy density compared with the
energy density. Both the non-relativistic limits in equation
(\ref{NL-limit}) and the subhorizon limit in equation
(\ref{SS-limit}) correspond to taking $c \rightarrow \infty$ limit.

To the linear order, the vector-type part of equation (\ref{eq3})
gives \bea
   c^2 {\Delta \over a^2} \Psi_i^{(v)}
       = - 16 \pi G \varrho {\widehat v_i^{(v)} \over c}.
\eea This is ``the initial value equation for the frame-dragging
potential $\Psi_i^{(v)}$'', see equation (4.12) in \cite{Bardeen-1980}; $\Psi_i^{(v)}$ is
supported by $\widehat v_i^{(v)}$ which is related to the fluid vorticity
\cite{Bardeen-1980}. Thus, we notice that in the subhorizon limit
$\Psi_i^{(v)}$ is suppressed by a factor in equation
(\ref{SS-limit}) compared with the $\widehat v_i^{(v)}/c$. Therefore, we can
ignore $\Psi_i^{(v)}$ in the non-relativistic limit while keeping $\widehat v_i^{(v)}$. Thus, we have \bea
   \chi_i = c \chi_{,i},
   \label{chi_i}
\eea

%
%
\subsection{Proof in the Zero-shear gauge}
                                          \label{sec:ZSG}

The zero-shear gauge sets $\chi \equiv 0$ \cite{Harrison-1967}. As
we have $\chi_i = c \chi_{,i}$ due to the reason presented above
equation (\ref{chi_i}), we have $\chi_i = 0$. The condition $\chi_i
= 0$ leads to vanishing shear of the normal-frame vector to the
fully nonlinear order \cite{Bardeen-1980,Hwang-Noh-2013}, thus we
term it as the zero-shear gauge; in the literature it is often known
as the conformal Newtonian gauge, the longitudinal gauge, or the
Poisson's gauge, with less clear reasons.

In the subhorizon limit together with the non-relativistic limits,
equations (\ref{eq3}) and (\ref{eq5}), respectively, give \bea
   & & \kappa = - {12 \pi G a \over c^2} \Delta^{-1}
       \nabla \cdot \left( \widetilde \varrho {\bf v} \right),
   \label{kappa-ZSG} \\
   & & \varphi = - \alpha.
   \label{varphi-ZSG}
\eea Using equations (\ref{kappa-ZSG}) and (\ref{varphi-ZSG}) and
using the Newtonian identifications in equation
(\ref{identification}) we can show the following. Equation
(\ref{eq6}) gives equation (\ref{mass-conservation}) Equation
(\ref{eq7}) gives equation (\ref{momentum-conservation}). Both
equations (\ref{eq2}) and (\ref{eq4}) give equation
(\ref{Poisson-eq}). Finally, equation (\ref{eq1}) gives equation
(\ref{U-eq}). Therefore, we proved the exact Newtonian
correspondence in the zero-shear gauge.

Equation (\ref{eq6}) is the energy conservation equation. Later in
section \ref{sec:E-conservation} we will show that to the leading
($c^0$, thus Newtonian) order we have equation
(\ref{mass-conservation}) which is the mass conservation, whereas,
to the next ($c^{-2}$, thus first post-Newtonian) order we have the
proper energy conservation equation.

%
%
\subsection{Proof in the Uniform-expansion gauge}
                                          \label{sec:UEG}

The uniform-expansion gauge sets $\kappa \equiv 0$; this was
introduced as the uniform-Hubble gauge in \cite{Bardeen-1980}.

In the subhorizon limit together with the non-relativistic limits,
equations (\ref{eq3}) and (\ref{eq5}), respectively, give \bea
   & & c \chi = - {12 \pi G a^3 \over c^2} \Delta^{-1}
       \nabla \cdot \left( \widetilde \varrho {\bf v} \right),
   \label{chi-UEG} \\
   & & \varphi = - \alpha.
   \label{varphi-UEG}
\eea Using equations (\ref{chi-UEG}) and (\ref{varphi-UEG}) and
using the Newtonian identifications in equation
(\ref{identification}) we can show the following. Equation
(\ref{eq6}) gives equation (\ref{mass-conservation}) which is the mass conservation equation; for the energy conservation part, see section \ref{sec:E-conservation}. Equation
(\ref{eq7}) gives equation (\ref{momentum-conservation}). Both
equations (\ref{eq2}) and (\ref{eq4}) give equation
(\ref{Poisson-eq}). Finally, equation (\ref{eq1}) gives equation
(\ref{U-eq}). Therefore, we proved the exact Newtonian
correspondence in the uniform-expansion gauge.

%
%
\section{Energy conservation equation as the post-Newtonian order}
                                                      \label{sec:E-conservation}

The referee has suggested us to show the energy conservation
equation in Newton's gravity, and here we present it. The energy
conservation equation follows from equation (\ref{eq6}) by keeping
the internal energy and the pressure terms to first post-Newtonian
(1PN) order. In both the zero-shear gauge and the uniform-expansion
gauge equation (\ref{eq6}) gives \bea
   \fl \left( 1 + {1 \over c^2} \widetilde \Pi \right)
       \left[ \dot {\widetilde \varrho}
       + 3 {\dot a \over a} \widetilde \varrho
       + {1 \over a} \nabla \cdot \left( \widetilde \varrho {\bf v} \right)
       \right]
       + {1 \over c^2} \widetilde \varrho \left[
       \dot {\widetilde \Pi}
       + {1 \over a} {\bf v} \cdot \nabla \widetilde \Pi
       + \left( 3 {\dot a \over a}
       + {1 \over a} \nabla \cdot {\bf v} \right)
       {\widetilde p \over \widetilde \varrho} \right]
       = 0.
   \label{E-conservation-PN}
\eea The first part gives the mass conservation in equation
(\ref{mass-conservation}) and the second part gives the energy
conservation equation \bea
   & & \dot {\widetilde \Pi}
       + {1 \over a} {\bf v} \cdot \nabla \widetilde \Pi
       + \left( 3 {\dot a \over a}
       + {1 \over a} \nabla \cdot {\bf v} \right)
       {\widetilde p \over \widetilde \varrho}
       = 0.
\eea Apparently the pure energy conservation part comes from the 1PN
order. The proper 1PN expansion gives the pure energy conservation part as the 1PN correction terms in the mass conservation equation;
see \cite{Chandrasekhar-1965} in the Minkowsky background, and equations
(57)-(63) in \cite{Hwang-etal-2008} which include the anisotropic
stress and flux.

In the covariant notation, the energy-conservation equation is \bea
   & & 0 = \widetilde u^a \widetilde T^b_{a;b}
       = \widetilde u^a
       \left\{ \left[ \widetilde \varrho \left( c^2 + \widetilde \Pi
       \right)
       + \widetilde p \right) \widetilde u^b \widetilde u_a
       + \widetilde p \delta^b_a \right\}_{;b}
   \nonumber \\
   & & \qquad
       = - \left( c^2 + \widetilde \Pi \right)
       \left( \widetilde \varrho \widetilde u^c \right)_{;c}
       - \widetilde \varrho
       \left( \widetilde \Pi_{,c} \widetilde u^c
       + \widetilde \theta \widetilde p/ \widetilde \varrho \right),
   \label{E-conservation-cov}
\eea where $\widetilde \theta \equiv \widetilde u^c_{\;\, ;c}$ is
the expansion scalar. The mass conservation demands $\left(
\widetilde \varrho \widetilde u^c \right)_{;c} = 0$ and the
remaining part of equation (\ref{E-conservation-cov}) gives the
energy conservation equation. Equation (\ref{E-conservation-cov})
leads to equation (\ref{E-conservation-PN}).

%
%
\section{Newtonian limit in Einstein's gravity}
                                                      \label{sec:Newtonian-limit}

Our proof of the Newtonian limit in the two fundamental gauge
conditions shows that, in order to have the Newtonian hydrodynamic
equations we may set $\chi_i = 0$ but should have $\varphi = -
\alpha$, thus the metric should be \cite{Chandrasekhar-1965} \bea
   ds^2 = - \left( 1 - {1 \over c^2} 2 U \right) c^2 d t^2
       + a^2 \left( 1 + {1 \over c^2} 2 U \right)
       \delta_{ij} d x^i d x^j.
   \label{metric-Newtonian}
\eea This may look strange because it is well known that the
Newtonian equations can be recovered from the Newtonian
gravitational potential contained in $\widetilde g_{00}$ only.
Whereas, our derivation apparently demanded the presence of
nontrivial $\widetilde g_{ij}$ as well; the $U$ term in $\widetilde
g_{ij}$ is the well-known post-Newtonian correction important in
affecting the light deflection, see equation (149) of
\cite{Hwang-etal-2008}. Here we would like to make a comment on this
subtle issue.

In order to have the conservation equations and the Poisson's
equation in (\ref{mass-conservation})-(\ref{Poisson-eq}) from
equations (\ref{eq6}), (\ref{eq7}) and (\ref{eq4}), we only need the
$U$ term in $\widetilde g_{00}$ part. Whereas, to get equations
(\ref{Poisson-eq}) and (\ref{U-eq}) from equations (\ref{eq2}) and
(\ref{eq1}), respectively, we require $\varphi = - \alpha$ which is
given by equation (\ref{eq5}), thus demanding the presence of $U$
term in $\widetilde g_{ij}$ part which is ordinarily the first
post-Newtonian contribution.

Therefore, the situation can be summarized as the following. The
complete Newtonian equations indeed follow from parts of Einstein's
equations based on $\widetilde g_{00}$ part only. Meanwhile, the
validity of the rest of Einstein's equations {\it demands} the
presence of nontrivial $\widetilde g_{ij}$ for the self consistency.
We can also interpret the situation as that the $\widetilde g_{00}$
part gives the Newtonian limit, and $\widetilde g_{ij}$ part is
simply determined by remaining parts of the Einstein's equation for
self consistency with the Newtonian limit; yet in another words,
parts of Einstein's equation mix the Newtonian and the
post-Newtonian orders
\cite{Chandrasekhar-1965,Kofman-Pogosyan-1995,Matarrese-Terranova-1996}.

%
%
\section{Discussion}
                                                      \label{sec:Discussion}

In this work we have {\it proved} that, in the subhorizon limit
together with the non-relativistic (weak gravity, slow-motion, and
negligible pressure and internal energy density compared with the
energy density) limits, Newtonian nonlinear perturbation equations
are exactly recovered from the general relativistic ones in the
zero-shear gauge and in the uniform-expansion gauge. As a
consequence, we also have shown that the Newtonian hydrodynamic
equations are exactly recovered in the Minkowsky background.

We point out that in our previous proof, based on the growing mode
solutions in the zero-pressure medium, of the correspondence to the
second order we {\it only} have assumed the subhorizon limit, thus
the correspondence is valid for relativistic (strong gravity and
fast motion) case \cite{Hwang-etal-2012}, whereas in our present
general proof valid to fully nonlinear order we additionally have
assumed the weak gravity as well as slow-motion limits.

Our proof of the relativistic/Newtonian correspondence to the fully
nonlinear order in the two gauges is not necessarily a trivial one.
For example, the correspondence is {\it not} available in other
fundamental gauge conditions, like the comoving gauge ($v \equiv
0$), the uniform-curvature gauge ($\varphi \equiv 0$),
uniform-density gauge ($\delta \equiv 0$), and the synchronous gauge
($\alpha \equiv 0$), even to the linear order in the subhorizon
limit \cite{Hwang-1994,Hwang-Noh-1999,Hwang-etal-2012}; notice that each one of
these other fundamental gauges sets the fundamental fluid (density
or velocity) or potential perturbation variables in Newtonian theory
equal to zero. Our result that the correspondence is possible only
in the two particular gauge conditions is consistent with the fact
that the post-Newtonian approach is consistent only with the
zero-shear gauge and the uniform-expansion gauge, see
\cite{Hwang-etal-2008,Noh-Hwang-2012}.

For the correspondences we have demanded the non-relativistic as
well as the sub-horizon limits. As we approach the horizon scale and
as soon as the (both special and general)
relativistic effects are not negligible, we have pure Einstein's
gravity correction terms appearing. In order to properly handle the
relativistic effects we may have three different but complimentary
approaches: the relativistic (nonlinear) perturbation theory, the
cosmological post-Newtonian approach, and the full-blown general
relativistic numerical relativity.

The relativistic/Newtonian correspondence in all cosmological scales
was studied up to the second-order perturbation in
\cite{Hwang-Noh-1999,Hwang-etal-2012}. The relativistic/post-Newtonian
correspondence 
was studied based on the linear perturbation theory and the first
post-Newtonian approximation in \cite{Noh-Hwang-2012}.
The results show that, for the correspondence, different gauges
are suitable for different variables: e.g., the comoving gauge shows
exact relativistic/Newtonian correspondences for the density and the
velocity perturbations up to the second order, whereas the zero-shear
gauge shows exact relativistic/Newtonian correspondence for the
gravitational potential perturbation to the linear order only \cite{Hwang-etal-2012}.

In this work we have considered the situation where the pressure
term is negligible compared with the energy density. Thus we ignored
any gravitating role of the pressure. The situation is applicable in
the matter dominated era in the $\Lambda$CDM cosmology; $\Lambda$ is
the cosmological constant and CDM indicates the cold dark matter.
However, as we consider other than $\Lambda$ as the dark energy or
other than CDM as the dominant dark matter, some form of pressure
term will become important, thus demanding situations deviating from
the Newtonian theory. The radiation pressure term could be important
in the early matter dominated era just after the radiation-dominated
era or the recombination. In these situations where the pressure
term plays the gravitating roles it is necessary that we should go
back to Einstein's gravity displayed in equations
(\ref{eq1})-(\ref{K-bar-eq}). Whether the gravitating role of
pressure can be handled by a minimal extension of Newtonian
equations is left for a future study.

%
%
\section*{Acknowledgments}

We wish to thank the referee for useful and important suggestions.
H.N.\ was supported by grant No.\ 2012 R1A1A2038497 from NRF. J.H.\ was supported by KRF Grant funded by the Korean Government (KRF-2008-341-C00022).

%
%



\section*{References}
\begin{thebibliography}{10}
\bibitem{Einstein-1917}
         A. Einstein, K\"oniglich Preussische Akademie der Wissenschaften (1917);
         translated in J. Bernstein and G. Feinberg, eds,
         {\it Cosmological Constants: Papers in Modern Cosmology} (Columbia Univ. Press, 1989).
\bibitem{Friedmann-1922}
         A.A. Friedmann, Zeitschrift f��ur Physik \textbf{10}, 377 (1922);
         translated in Gen. Rel. Grav. \textbf{31}, 1991 (1999).
\bibitem{Milne-McCrea-1934}
         E.A. Milne, Quart. J. Math. \textbf{5}, 64 (1934);
         W.H. McCrea and E.A. Milne, Quart. J. Math. \textbf{5}, 73 (1934).
\bibitem{Lifshitz-1946}
         E.M. Lifshitz, J. Phys. (USSR) \textbf{10}, 116 (1946).
\bibitem{Bonnor-1957}
         W.B. Bonnor, Mon. Not. R. Astron. Soc. \textbf{117}, 104 (1957).
\bibitem{Nariai-1969}
         H. Nariai, Prog. Theor. Phys., \textbf{41}, 686 (1969).
\bibitem{Layzer-1954}
         D. Layzer, Astron. J. \textbf{59}, 268 (1954);
         D.S. Lemons, Am. J. Phys. \textbf{56}, 502 (1988).
\bibitem{Hwang-etal-2008}
         J. Hwang, H. Noh and D. Puetzfeld, JCAP \textbf{03}, 010 (2008).
\bibitem{Chandrasekhar-1965}
         S. Chandrasekhar, Astrophys. J. \textbf{142}, 1488 (1965).
\bibitem{Bardeen-1980}
         J.M. Bardeen, Phys. Rev. D \textbf{22}, 1882 (1980).
\bibitem{Hwang-1994}
         J. Hwang, Astrophys. J. \textbf{427}, 533 (1994).
\bibitem{Hwang-Noh-1999}
         J. Hwang and H. Noh, Gen. Rel. Grav. \textbf{31}, 1131 (1999).
\bibitem{Hwang-etal-2012}
         J. Hwang, H. Noh and J. Gong, Astrophys. J. \textbf{752}, 50 (2012).
\bibitem{Hwang-Noh-2013}
         J. Hwang and H. Noh, Mon. Not. R. Astron. Soc. submitted, arXiv:1207.0264v2 (2013).
\bibitem{Bardeen-1988}
         J.M. Bardeen, {\it Particle Physics and Cosmology}, edited by
                       L. Fang and A. Zee (Gordon and Breach, London, 1988) 1;
         J. Hwang, Astrophys. J. \textbf{375}, 443 (1991).
\bibitem{Peebles-1980}
         P.J.E. Peebles, {\it The large-scale structure of the universe},
                (Princeton Univ. Press, Princeton, 1980).
\bibitem{Harrison-1967}
         E.R. Harrison, Rev. Mod. Phys. \textbf{39}, 862 (1967).
\bibitem{Kofman-Pogosyan-1995}
         L. Kofman and D. Pogosyan, Astrophys. J. \textbf{442}, 30 (1995).
\bibitem{Matarrese-Terranova-1996}
         S. Matarrese and D. Terranova, Mon. Not. R. Astron. Soc. \textbf{283}, 400 (1996).
\bibitem{Noh-Hwang-2012}
         H. Noh and J. Hwang, Astrophys. J. \textbf{757}, 145 (2012).
\end{thebibliography}
\end{document}